\documentstyle[aps,prl,preprint,epsfig]{revtex}
\begin{document}
\draft

\preprint{\tighten\vbox{\hbox{\hfil BIHEP-EP1-2000-02}}}

\title{Partial Wave Analysis of $J/\psi \to \gamma (K^{\pm }K_S^0\pi^{\mp})$}

\author{
J.~Z.~Bai,$^{1}$    Y.~Ban,$^{6}$      J.~G.~Bian,$^{1}$  A.~D.~Chen,$^{1}$
G.~P.~Chen,$^{1}$   H.~F.~Chen,$^{2}$  H.~S.~Chen,$^{1}$  J.~C.~Chen,$^{1}$
X.~D.~Chen,$^{1}$   Y.~Chen,$^{1}$
% ...
\\(BES Collaboration)\cite{besjpsi}\\
D.~V.~Bugg $^{12}$ and B.~S.~Zou $^{1,12}$}

\vspace{1cm}

\address{
$^{1}$ Institute of High Energy Physics, Beijing 100039,
People's Republic of China \\
$^{2}$ University of Science and Technology of China, Hefei 230026,
People's Republic of China \\
$^{3}$ Shandong University, Jinan 250100,
People's Republic of China \\
$^{4}$ Hangzhou University, Hanzhou 310028,
People's Republic of China \\
$^{5}$ Shanghai Jiaotong University, Shanghai 200030,
People's Republic of China \\
$^{6}$ Peking University, Beijing 100871,
People's Republic of China \\
$^{7}$ Hua Zhong Normal University, Wuhan 430079,
People's Republic of China \\
$^{8}$ China Center for Advanced Science and Technology(CCAST), World
Laboratory, Beijing 100080, People's Republic of China)\\
$^{9}$ Nankai University, Tianjin 300071,
People's Republic of China \\
$^{10}$ Henan Normal University, Xinxiang 453002,
People's Republic of China \\
$^{11}$ Wuhan University, Wuhan 430072,
People's Republic of China \\
$^{12}$ Queen Mary and Westfield College, London E1 4NS, United Kingdom}

\date{\today}

\maketitle

\begin{abstract}
BES data on $J/\psi \to \gamma (K^{\pm }K_S^0\pi^{\mp })$ are presented.
There is a strong peak due to $\eta (1440)/\iota$, which is fitted with
a Breit-Wigner amplitude with $s$-dependent widths for decays to
$K^*K$, $\kappa K$, $\eta \pi \pi$ and $\rho \rho$; $\kappa$ refers
to the $K\pi$ S-wave.
At a $K\bar{K}\pi$ mass of $\sim 2040$ MeV, there is a second peak with width
$\sim 400$ MeV; $J^P = 0^-$ is preferred over $1^+$ and $2^-$ respectively 
by 5.2 and 6.8 standard deviations.
It is a possible candidate for a $0^-$ $s\bar sg$ hybrid partner of $\pi
(1800)$.
\end{abstract}

\vspace{0.5cm}
\pacs{PACS numbers: 14.40.Cs, 12.39.Mk, 13.25.Jx, 13.40.Hq}

There have been earlier data from Mark III \cite{1} and DM2 \cite{2} 
for $J/\psi$ radiative decays to $K^{\pm }K_S^0\pi^{\mp }$, as well as
$K^+K^-\pi ^0$.
Recently, the BES group has published data on the latter channel \cite{3}.
Here we present BES data on decays to $K^{\pm }K_S^0\pi^{\mp}$.
These data have lower backgrounds than for $K^+K^-\pi ^0$,
because of the identification of $K_S^0 \to \pi ^+\pi ^-$.
Consequently, the partial wave analysis may be extended up to a
$K\bar{K}\pi$ mass of 2300 MeV, covering an interesting structure
at $\sim 2040$ MeV.

The Beijing Spectrometer(BES) has collected $7.8 \times
10^6 ~J/\psi$ triggers, used here.
Details of the detector are given in Ref. \cite{4}. 
We describe briefly those detector elements playing a crucial role in 
the present measurement. 
Tracking is provided by a 10 superlayer main drift chamber (MDC). 
Each superlayer contains four layers of sense wires measuring both
the position and the ionization energy loss ($d$E$/dx$) of charged particles. 
The momentum resolution is $\sigma_P/P = 1.7\%\sqrt{1 + P^2}$,
where $P$ is the momentum of charged tracks in GeV/$c$. 
The resolution of the $d$E$/dx$ measurement is $\sim \pm 9\%$,  
providing good $\pi/K$ separation and proton identification for momenta
up to 600 MeV/c. 
An array of 48 scintillation counters surrounding the MDC measures the
time-of-flight (TOF) of charged tracks with a resolution of 330$ps$ for 
hadrons. 
Outside the TOF system is
an electromagnetic calorimeter made of lead sheets and streamer tubes and
having a $z$ positional resolution of 4 cm. 
The energy resolution scales as 
$\sigma_E/E = 22\%/\sqrt{E}$, where $E$ is the energy in GeV. 
Outside the shower counter is a solenoidal magnet producing a 0.4 Tesla
magnetic field.

Each candidate event is required to have four charged tracks.
Each track must have a good helix fit in the polar angle
range $-0.8 < \cos\theta < 0.8$ and a transverse momentum $>60$ MeV/c.
A vertex is required within an interaction region $\pm 30$ cm longitudinally
and 3 cm radially. A positive identification of just one $K^{\pm}$ is
required using time of flight and/or $dE/dx$.
Events are fitted kinematically to the 4C hypothesis
$J/\psi \to \gamma (K^{\pm}\pi^{\mp}\pi^+\pi^-)$, 
requiring a confidence level $>5\%$. 

Backgrounds arise mainly from $\pi^0 K^{\pm} \pi^{\mp} \pi^+ \pi^-$ and 
$K^{\pm} \pi^{\mp} \pi^+ \pi^-$.
Those events giving a better fit to these channels are rejected.
Next, we require
$\mid{U_{miss}}\mid = \mid E_{miss}-P_{miss} \mid <0.15$ GeV/c$^2$,
so as to reject the events
with multi-photons or more or less than one charged kaon; 
here, $E_{miss}$ and $P_{miss}$ are,
respectively, the missing energy and missing momentum of all charged particles.
The momentum of the $K^{\pm} \pi^{\mp} \pi^{+} \pi^{-}$ 
system transverse to the photon 
$P_{t\gamma}^2=4\mid{P_{miss}}\mid{^2}~\sin^2(\theta_{m\gamma}/2) 
<0.005$ (GeV/c)$^2$ is required in order to remove the background 
$J/\psi\rightarrow
\pi^0K^{\pm}\pi^{\mp}\pi^{+}\pi^{-}$; here $\theta_{m\gamma}$ is the angle
between the missing momentum and the photon direction. 
Finally, $K_s^0$ are selected with a cut on the $\pi^+\pi^-$ invariant mass,
$\mid M_{\pi^+\pi^-}-M_{K_s^0} \mid < 25$ MeV.
Fig.~\ref{re-bg}(a) shows the $\pi^+\pi^-$ invariant mass
closest to the $K_S^0$ mass before the $K_s^0$ are selected;
a very strong signal $K_S^0$ is seen.
The number of surviving events is 1095 with 57 $\pm$ 5 non-$K_S^0$
background under the $K_S^0$.
For our final fit, we use 683 events below a $K\bar{K}\pi$ mass of 2.3 GeV.
A constraint to the $K_S^0$ vertex does not improve the signal/background
ratio further, but loses some events.


The effects of the various selection cuts on the data is simulated with
a full Monte Carlo of the BES detector including the decay path of the 
$K_s^0$; 250,000 Monte Carlo events are successfully fitted to
$J/\psi \to \gamma (K^{\pm }K_S^0\pi^{\mp})$. All background reactions
are similarly fitted to this channel. 
The estimated background is $29 \pm 7\%$, mostly from 
$J/\psi \to \pi ^0 (K^{\pm }K_S^0\pi^{\mp})$, some from non-$K_S^0$ events.
It peaks at about 2.3 GeV, and follows phase space closely.
We have included it in the amplitude analysis,
but it has little effect, since all genuine signals have a characteristic 
dependence on either or both of production and decay angles. 

Fig.~\ref{re-bg}(b) shows the $K^{\pm}K_S^0\pi^{\mp}$ mass spectrum; the dark
histogram shows the estimated background in the analysis region.
There is a conspicuous and somewhat asymmetric peak due to $\eta (1440)/\iota$,
similar to the earlier data from Mark III, DM2 and BES.
At high mass, there is a distinct peak at 2040 MeV. 
Fig.~\ref{DPcomp} shows Dalitz plots for three mass ranges: 
(a) 1360-1560 MeV, 
(b) 1600--1750 MeV, and (c) 1800--2200 MeV; fits are shown in (d),
(e) and (f).
There is a conspicuous $K^*K$ decay mode in the first region of the $\eta
(1440)$. 
At higher masses, it disappears rapidly, and the mass projections shown
in Fig.~\ref{proj_3} are consistent with decays to $\kappa K$ only;
above 1560 MeV, there is no  significant evidence for $a_0(980)K$,
$a_0 \to K\bar{K}$.

We have carried out a partial wave analysis using
amplitudes constructed from Lorentz-invariant
combinations of the 4-vectors and the photon polarization for
$J/\psi$ initial states with helicity $\pm 1$. Cross sections are summed
over photon polarisations. The relative magnitudes and phases of the 
amplitudes are determined by a maximum likelihood fit. 
We include $K\bar{K}\pi$ states with quantum numbers
$0^-$, $1^+$, $2^-$ and $2^+$.
There are two helicity amplitudes for $1^+$, three for $2^-$ and three
for $2^+$.
Because production is via an electromagnetic transition, 
the same phase is used for different helicity amplitudes to the same final
state. Different phases are allowed for different decay channels, e.g
$K^*K$ and $\kappa K$, because of strong interaction effects due to
rescattering.

The analysis is discussed separately for the mass region of $\eta (1440)$
and the 2040 MeV peak. 
The $\eta(1440)$ has been fitted using a Breit-Wigner amplitude with
$s$-dependent width:
\begin {equation}
f = \frac {\Lambda}{M^2 - s - iM[\Gamma _{K^*K}(s) + \Gamma _{\eta \sigma }(s)
+ \Gamma _{\rho \rho }(s) + \Gamma _{\kappa K}(s)]}.
\end{equation}
The numerator $\Lambda$ is a complex coupling constant. 
The $\Gamma(s)$ are taken to be proportional to the available phase space for
each channel, evaluated numerically \cite{5}. 
The $\eta \pi \pi$ phase space is taken from $\eta \sigma$, the dominant
channel, but $a_0(980)\pi$ phase space is similar and both are slowly varying
over this mass region. 
The magnitude of each $\Gamma$ is
adjusted iteratively so that cross sections integrated over the resonance
agree with the branching ratios determined experimentally.
The magnitude of $\Gamma _{\eta \sigma}$ has been obtained from BES data on
radiative decays to $\eta \pi \pi$ \cite{6}.
That for $\Gamma _{\rho \rho}$ has been obtained by fitting BES 
data on radiative decays to $4\pi$ \cite{7},
including in the fit $\eta (1440)$ and the broad $\eta(1800)$.
Here $\eta (1800)$ refers to the very broad  
$0^-$ signal ($\Gamma \simeq 1$ GeV) derived by Bugg and Zou \cite{8}
from an analysis of several channels of $J/\psi$ radiative decay.
Values of $\Gamma _{K^*K}$ and $\Gamma _{\kappa K}$ are obtained from
the present data.

In the mass region of the $\eta(1440)$, half the $\kappa K$ signal comes from
the low mass tail of $\eta(1800)$ and its constructive interference with
$\eta(1440)$. That is, if $\eta (1800)$ is removed from the fit,
the $\kappa K$ width of $\eta(1440)$ needs to be doubled.
Removing the $\eta(1800)$ has a significant, but not dramatic, effect on log
likelihood, which changes by 4.8 for 2 extra parameters.
   
The $f_1(1420)$ is also included in the amplitude analysis, and a small
component due to $f_1(1285)$. Both optimise close to the masses and widths
quoted by the Particle Data Group (PDG) \cite{9}, so we fix them at PDG values.
The amplitude analysis distinguishes cleanly between quantum numbers $1^+$ and
$0^-$ for $K^* K$ decays. If the whole $\eta (1440)$ signal is fitted with
$J^P = 1^+$ (optimising its mass and width), log likelihood is worse by 11.4, 
a significant amount. (Our definition of log likelihood is such that 
it increases by 0.5 for a one standard deviation change in one parameter). 

In the earlier analysis of BES data on the $K^+K^-\pi^0$ final state \cite{3},
a fairly large amplitude was fitted for $\eta(1800)$. The smaller
background in present data and the wider mass range allow us to
show that this component should in fact be rather small. Its effects on the
$\eta(1440)$ may be replaced with some increase in the total width
of that resonance and an increase in its width for decays to $K^*K$.
Present results for the fitted widths are shown in Fig.~\ref{iwidths} 
and branching fractions in Table 1.
The fit is compared with the $K\bar{K}\pi$ mass spectrum by the histogram in
Fig. 5. 

A free fit to the mass gives 1440 MeV. 
However, $\eta \pi \pi $ data give a resonance mass of $1405 \pm 5$ MeV, 
according to the summary by the Particle Data Group \cite{9}.
The $s$-dependent width we use for $\eta(1440)$ explains naturally a
mass difference of 20 MeV between $\eta \pi \pi$ and $K\bar{K}\pi$ data; 
the rapidly increasing phase space for $K^*K$ makes the $K\bar{K}\pi$
channel peak higher and explains also the asymmetric shape
of the peak, which rises rapidly on the lower side of the peak and falls
more slowly on the upper side.  A small ($\sim 15$ MeV) discrepancy remains
between the peaks fitted to $\eta \pi \pi$ and $K\bar{K}\pi$.
We adopt a compromise between fitting these data and $\eta \pi \pi$ by using
a mass of 1432 MeV, but the effect on other conclusions is negligible.
Interferences between $\eta (1440)$ and the broad $\eta (1800)$ depend on
their relative phases and can shift the peak in different data sets; 
so we do not regard this small discrepancy as a matter for concern.

Around 1650 MeV, there is some indication for a narrow $K\bar K\pi$ peak.
However, fitting it requires an unreasonably narrow width $\sim 30$ MeV.
An $s\bar s$ state at this mass has no obvious non-strange partners. 
If fitted, it is only
a two standard deviation effect. Therefore we discard it as a statistical
flutuation. Including it has negligible effects on parameters fitted to
$\eta (1440)$ and the peak at 2040 MeV.

We now turn to the latter peak.
It cannot be explained by the very broad $\eta(1800)$, which has a 
completely different and much flatter shape, illustrated by the shaded area in 
Figure~\ref{comps}(b) below.
We fit it with a simple Breit-Wigner amplitude of constant width. 
Its mass and width optimise at $M = 2040\pm50$ MeV, $\Gamma = 400\pm90$ MeV. 
We have tried fits to this peak with resonances having quantum numbers
$0^-$, $1^+$ and $2^-$; for standard $q\bar q$ states, one does not expect 
$3^+$ in kaonic channels until 2300 MeV. 
We find that log likelihood is better for $0^-$ than $1^+$ by 16.4. 
The latter has one additional parameter, so it is a poorer fit by 5.2 standard
deviations.  If a combination of $0^-$ and $1^+$ amplitudes is used,
log likelihood improves only by 0.6, and the fitted $1^+$ component
is very small: 4.4\% of $0^-$ in cross section. 
These results are not sensitive to the $\eta (1800)$ contribution: 
removing it, the distinction between 
quantum numbers $0^-$ and $1^+$ for the 2040 MeV peak remains at a log
likelihood difference of 12.9.

We have also tried adding or substituting $2^-$. 
Alone it gives a poor fit, worse in log likelihood than $0^-$ by 27.9. 
This demonstrates that $2^-$ and $0^-$ are well separated by their
distinctively different angular distributions. 
If it is added freely to the fit, it improves log likelihood by 3.2
for three extra parameters; this cannot be considered significant. 
Fig.~\ref{comps} shows magnitudes of components fitted in the amplitude
analysis when the 2040 MeV peak is fitted as $0^-$.
The slight differences between Figs.5(b) and (d) is due to
interferences of $\eta (1440)$ and $\eta (2040)$ with the broad
$\eta (1800)$ in Fig. 5(b).

Branching fractions for production and decay, including the dominant 
interferences, are given in Table 2. Values are integrated up to a 
$K\bar{K}\pi$ mass of 2.3 GeV. 
Decays to $K ^{\pm }K_S\pi^{\mp}$ have a branching ratio 1/3 of all $K\bar
K\pi$ decays.
We correct all measured branching ratios by this factor 3, so as to quote
branching fractions for all $K\bar K \pi$ charge states.
The overall branching fraction, summed over 
all final states is $(6.0 \pm 0.4 \pm 2.1)\times 10^{-3}$.

We now discuss possible interpretations for the 2040 MeV peak.
Our data for $J/\psi $ radiative decays to $\eta \pi ^+ \pi ^-$ [6] were 
fitted using an $\eta(1760)$ with a width of 250 MeV and an $\eta _2(1840)$.
The $\eta (1760)$ is entirely distinct from $\eta (1800)$, which has a
much larger width. A possible interpretation is that it is
the $n = 3~q\bar q$ state. 
Then the $\eta (2040)$ observed here could be its $s\bar s$ partner.

However, the VES collaboration has identified a $\pi(1800)$ \cite{10} 
with curious decay modes to $f_0(1300)\pi$, $f_0(980)\pi$ and $K_0(1430)K$,
but not $\rho \pi$.
There has been  speculation that this is an $I = 1$ hybrid \cite{11}.
The $\eta (1760)$ would make a natural partner; its decays to $\eta \sigma$
and $a_0(980)\pi$ are to be expected for a hybrid.
It is natural to expect a corresponding $s\bar sg$ state decaying to 
$\kappa K$ in the $K\bar{K}\pi$ channel roughly
200--250 MeV above the peak in $\eta \pi \pi$.
In $J/\psi$ radiative decays, the amplitude for production of 
$q\bar q$ states is suppressed by two powers of $\alpha _s$, required to 
couple intermediate gluons to quarks; at 2040 MeV, $\alpha _s \simeq 0.41$. 
Production of a hybrid will only be suppressed by one power of $\alpha _s$
in amplitude.
We therefore examine the possible interpretation of $\eta(1760)$ as a 
$q\bar q g$ hybrid.

For a hybrid, the branching fraction expected in the $K\bar{K}\pi$ channel is
half that for $\eta \pi \pi$, since in $J/\psi$ decays intermediate gluons
couple equally to $u\bar u$, $d\bar d$ and $s\bar s$. 
If fitted as $0^-$, the branching ratio for the 2040 MeV peak in $
K\bar K\pi$ is $(2.1 \pm 0.1 \pm 0.7)\times 10^{-3}$;
this value is obtained after allowing for interferences with
$\eta (1800)$ and includes the error in the overall normalisation.
It is to be compared with the branching ratio for $\eta (1760)$ in
$\eta\pi\pi$ of $(1.8 \pm 0.75) \times 10^{-3}$ \cite{6}.
These values are consistent within the sizable errors with the expectation 
for hybrids.

The magnitude of branching ratio we now fit to $\eta(1800) \to \kappa K$ is
$(0.58 \pm 0.03 \pm 0.20)\times 10^{-3}$. Again, the error includes the
overall normalisation uncertainty. It compares with
$(1.08 \pm 0.45)\times 10^{-3}$ fitted to $\eta \pi \pi$ decays \cite{6}.
Within the errors, these values are now consistent with flavour-blind
decays of a glueball. 

In summary, present data contain less background than earlier data on
$J/\psi \to \gamma (K^+K^-\pi ^0)$ and allow a somewhat improved
determination of the properties of $\eta (1440)$. Its dominant decay mode is
to $K^*K$. This suggests it is the first radial excitation of
$\eta(958)$, probably mixed with the broad $\eta(1800)$, in order to 
account for its strong production in $J/\psi$ radiative decays. 
We now find a small component of $\eta(1800)$ decaying to $\kappa K$.

We observe a peak at 2040 MeV which may be fitted with a $0^-$ 
resonance of width 400 MeV. 
$J^P = 0^-$ is preferred over $1^+$ and $2^-$ respectively 
by 5.2 and 6.8 standard deviations.
Its branching fraction, when compared with the $\eta \pi \pi$ channel,
would be  consistent with interpretation as a $0^-$ $s\bar sg$ hybrid. 

{\vspace{0.9cm}
The BES group thanks the staff of IHEP for technical support in running the
experiment. This work is supported in part by China Postdoctoral Science
Foundation and National Natural Science Foundation of China under contract
Nos. 19991480, 19825116 and 19605007;
and by the Chinese Academy of Sciences under contract No. KJ 95T-03(IHEP).
We also acknowledge financial support from the Royal Society
for collaboration between Chinese and UK groups.

\begin{center}

\begin{figure}[htbp]
\centerline{\epsfig{file=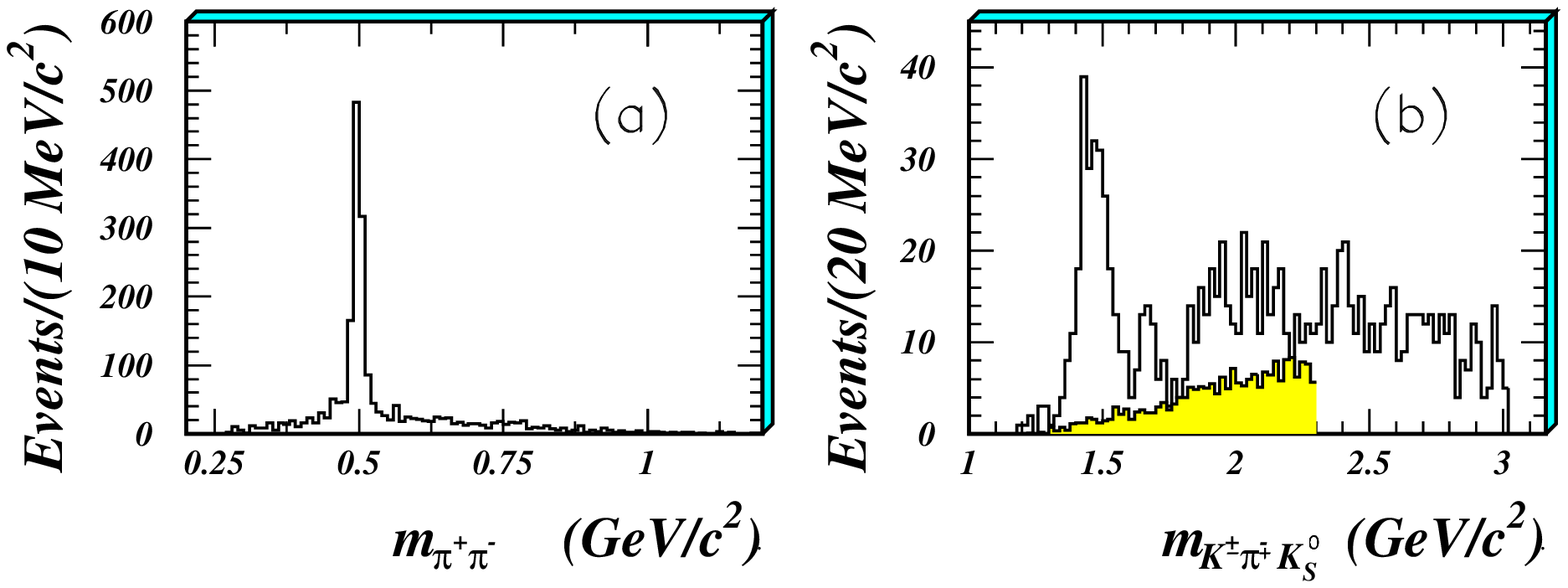,width=5.8in}} 
\caption[]{ (a) the $\pi^+\pi^-$ invariant mass with invariant mass
closest to the $K_S^0$ mass; (b) $K\bar{K}\pi$ mass spectrum.
The dark dashed histogram of (b) shows the estimated background
in the analysis region ($K\bar{K}\pi$ mass below 2.3 GeV). }
\label{re-bg}
\end{figure}

\begin{figure}[htbp]
\centerline{\epsfig{file=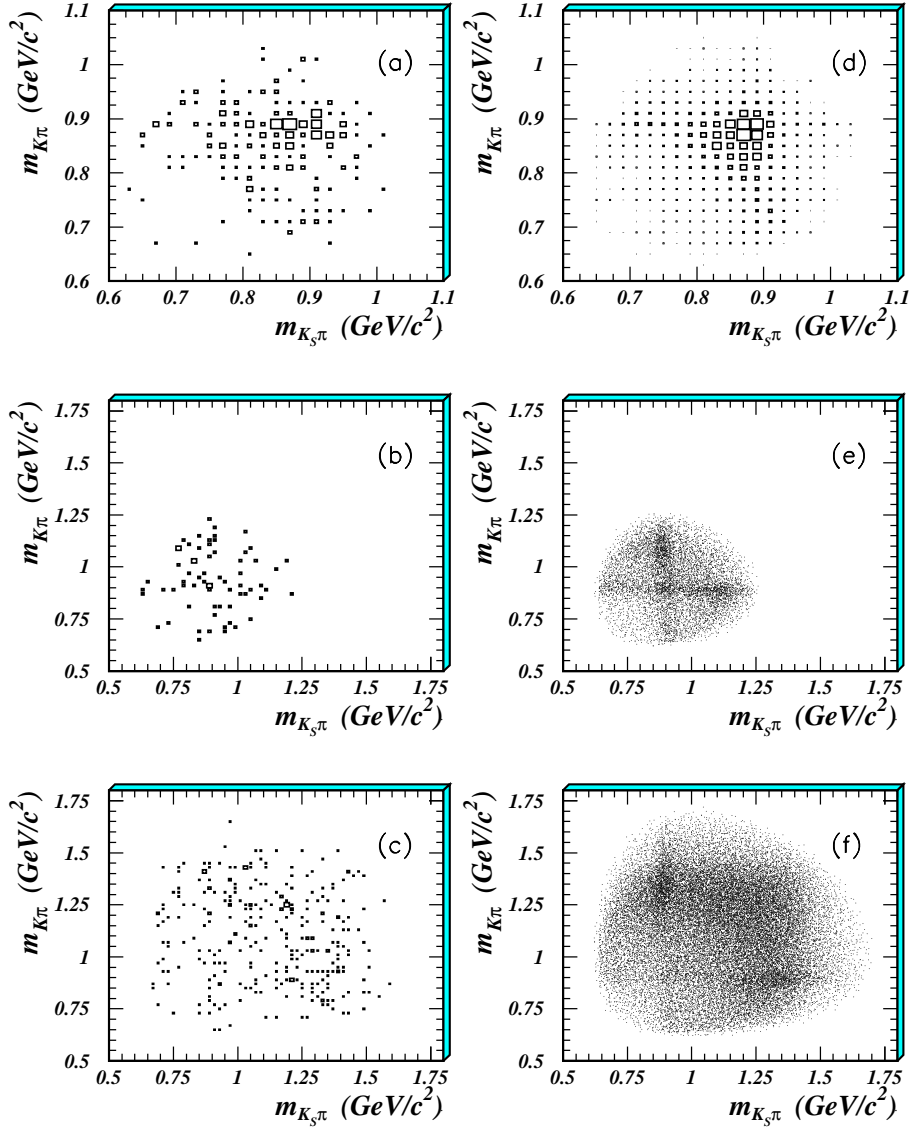,width=5.0in}} 
\caption[]{ Dalitz plots for $KK\pi$ mass ranges (a) 1360--1560 MeV,
(b) 1600--1750 MeV, (c) 1800--2200 MeV; (d), (e) and (f) show fitted
Dalitz plots.}
\label{DPcomp}
\end{figure}

\begin{figure}[htbp]
\centerline{\epsfig{file=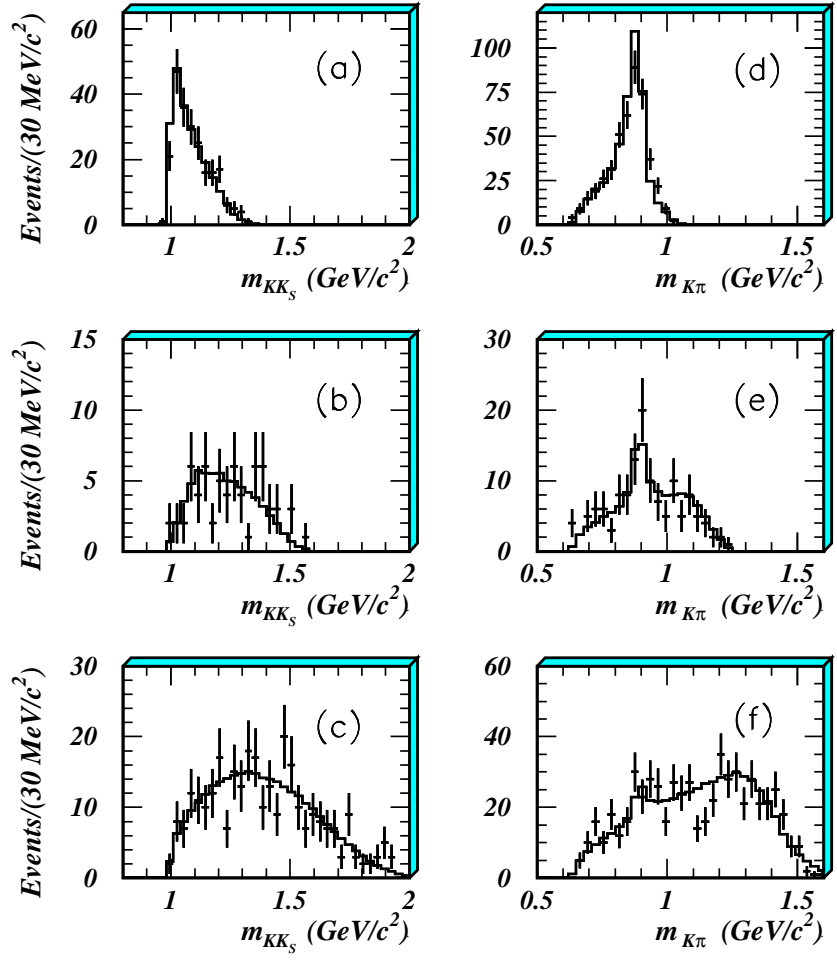,width=5.0in}} 
\caption[]{ Projections on to (a)--(c) $K^\pm K^0_S$ mass, (d)--(f) $K\pi$ mass for the
three mass intervals of Fig. 2; histograms show the fit. }
\label{proj_3}
\end{figure}

\begin{figure}[htbp]
\centerline{\epsfig{file=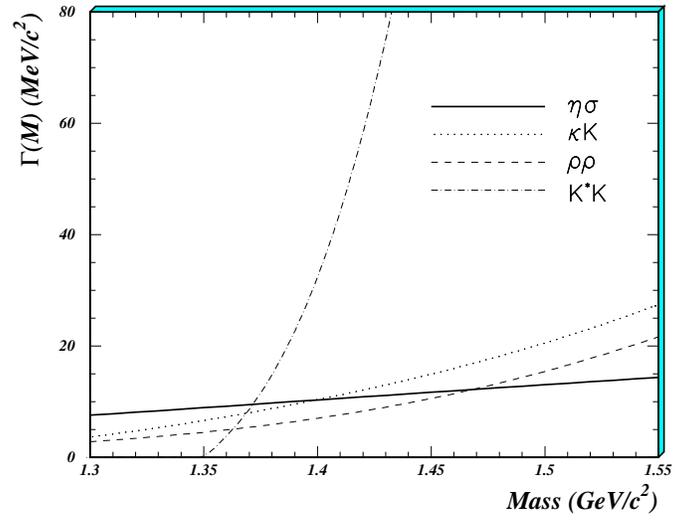,width=3.5in}} 
\caption[]{ The $s$-dependence of widths fitted to $\eta (1440)$.}
\label{iwidths}
\end{figure}

\newpage
\begin{figure}[htbp]
\centerline{\epsfig{file=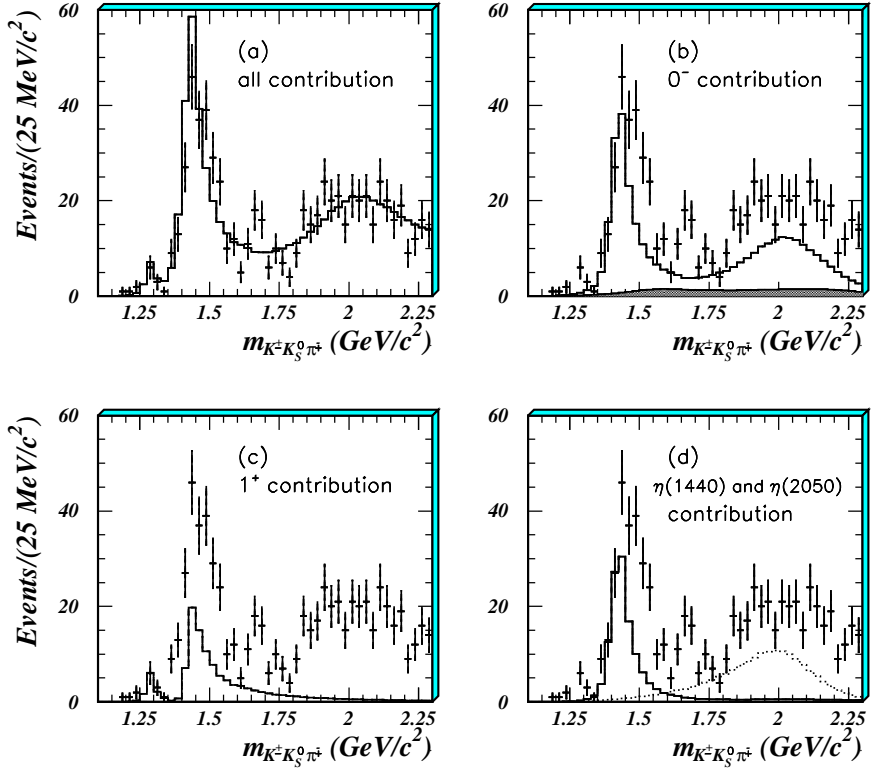,width=5.0in}} 
\caption[]{ Projections from all events below a $K ^{\pm }K_S\pi^{\mp}$ 
mass of 2.3 GeV of
(a) all contributions, (b) $0^-$ including interferences, (c) $1^+$ and 
(d) $\eta (1440)$ (full curve) and 
$\eta(2040)$ (dotted) without interferences; 
the dark shaded histogram of (b) is the contribution
of $\eta(1800)$. Crosses are data and histograms the fit. }
\label{comps}
\end{figure}

\begin{table}[htp]
\caption {Branching ratios (BR) of $\eta (1440)$ integrated over its width.}
\begin{center}
\begin{tabular}{|cc|} 
\hline                                       
Decay Channel & BR (\%) \\\hline
$K^*K$ & $0.70 \pm 0.05$ \\
$\kappa K$ & $0.13 \pm 0.03$ \\
$\eta \pi \pi$ & $0.09 \pm 0.03$ \\
$\rho \rho$ & $0.08 \pm 0.03$\\\hline
\end {tabular}
\end{center}
\end{table}

\begin{table}[htp]
\caption {Branching fractions (BF) for production and decay. 
Values are corrected for all charge states in $K\bar{K}\pi$.}
\begin{center}
\begin{tabular}{|lc|} 
\hline                                       
Process  & BF (\%) \\\hline
1) $J/\psi \to \gamma \eta (1440)$, $\eta (1440) \to K\bar{K}\pi$  & 
$(1.66 \pm 0.10 \pm 0.58)\times 10^{-3}$ \\
2) $J/\psi \to \gamma f_1 (1285)$, $f_1 (1285) \to K\bar{K}\pi$  & 
$(0.61 \pm 0.04 \pm 0.21)\times 10^{-3}$ \\
3) $J/\psi \to \gamma f_1 (1420)$, $f_1 (1420) \to K\bar{K}\pi$  & 
$(0.68 \pm 0.04 \pm 0.24)\times 10^{-3}$ \\
4) $J/\psi \to \gamma \eta (1800)$, $\eta (1800) \to \kappa K$  & 
$(0.58 \pm 0.03 \pm 0.20)\times 10^{-3}$  \\
5) Interference between (1) and (4) & $(0.15 \pm 0.01 \pm 0.05)\times 10^{-3}$ \\
6) Interference between (1) (3)  & $(-0.03 \pm 0.01 \pm 0.01)\times 10^{-3}$ \\
7)$J/\psi \to \gamma \eta (2040)$, $\eta (2040) \to  \kappa K$ & 
$(2.1 \pm 0.1 \pm 0.7)\times 10^{-3}$  \\\hline 
\end {tabular}
\end{center}
\end{table}

\end{center}


\begin{thebibliography}{99}

\bibitem[\dag]{besjpsi} Data analyzed were taken prior to the participation
                  of U.S. members of the BES Collaboration.

\bibitem{1}  J.Z. Bai et al., Phys. Rev. Lett. 65 (1990) 2507.
\bibitem{2}  J.-E. Augustin et al., Phys. Rev D46 (1992) 1951.
\bibitem{3}  J.Z. Bai et al., Phys. Lett. B440 (1998) 217.
\bibitem{4}  BES Collaboration, Nucl. Instr. Methods, {\bf A344} (1994) 319.
\bibitem{5}  D.V. Bugg, A.V. Sarantsev and B.S. Zou, Nucl. Phys. B471 (1996)
             59, equn. (40).
\bibitem{6}  J.Z. Bai et al., Phys. Lett. B446 (1999) 356.
\bibitem{7}  J.Z. Bai et al., Phys. Lett. B472 (2000) 207.
\bibitem{8}  D.V.Bugg and B.S.Zou, Phys. Lett. B 396 (1997) 295.
\bibitem{9}  Particle Data Group, Euro. Phys. J. C3 (1998) 1. 
\bibitem{10} D. Amelin et al., Phys. Lett.  B356 (1995) 595.
\bibitem{11} F.E. Close and P.R. Page, Nucl. Phys. B443 (1995) 233 and 
             Phys. Rev.  D52 (1995) 1706. 
\bibitem{12} J. Adomeit  et al., Zeit. Phys. C71 (1996) 227.
\bibitem{13} J.Z. Bai et al., Phys. Lett. B472 (2000) 200.

\end{thebibliography}
\end{document}